\documentclass[twocolumn,prb,showpacs,amsmath,amssymb,superscriptaddress]{revtex4}
\usepackage{graphicx,bm,dcolumn,amsmath,amssymb,amsfonts}
\newcommand{\Hop}{\hat{H}}

\newcommand{\psiop}{\hat{\psi}}
\newcommand{\vecr}{\bm{r}}
\newcommand{\veck}{\bm{k}}

\newcommand{\vecq}{\bm{q}}
\newcommand{\piD}{\mathcal{D}}
\newcommand{\Tr}{{\rm Tr}\,}

\newcommand{\Gop}{\hat{G}}

\newcommand{\Kop}{\hat{K}}
\newcommand{\Iop}{\hat{I}}

\newcommand{\Uop}{\hat{U}}

\newcommand{\GVec}[1]{\mbox{\boldmath$#1$}}

\begin{document}

\title{Collective modes in multiband superfluids and superconductors: \\Multiple dynamical classes}
\affiliation{
CCSE, Japan Atomic Energy Agency, 
Higashi-Ueno, Tokyo 110-0015, Japan}
\affiliation{
CREST(JST), Honcho, Kawaguchi, Saitama 332-0012, Japan}
\affiliation{
JST, TRIP, Sanbancho, Tokyo 102-0075, Japan}
\affiliation{
Institute for Materials Research, Tohoku University, Katahira, Sendai
980-8577, Japan} 
\affiliation{
Department of Physics, University of Tokyo, Hongo, Tokyo 113-0033, Japan}
\author{Yukihiro Ota}
\affiliation{
CCSE, Japan Atomic Energy Agency, 
Higashi-Ueno, Tokyo 110-0015, Japan}
\affiliation{
CREST(JST), Honcho, Kawaguchi, Saitama 332-0012, Japan}
\author{Masahiko Machida}
\affiliation{
CCSE, Japan Atomic Energy Agency, 
Higashi-Ueno, Tokyo 110-0015, Japan}
\affiliation{
CREST(JST), Honcho, Kawaguchi, Saitama 332-0012, Japan}
\affiliation{
JST, TRIP, Sanbancho, Tokyo 102-0075, Japan}
\author{Tomio Koyama}
\affiliation{
CREST(JST), Honcho, Kawaguchi, Saitama 332-0012, Japan}
\affiliation{
Institute for Materials Research, Tohoku University, Katahira, Sendai 980-8577, Japan} 
\author{Hideo Aoki}
\affiliation{
JST, TRIP, Sanbancho, Tokyo 102-0075, Japan}
\affiliation{
Department of Physics, University of Tokyo, Hongo, Tokyo 113-0033, Japan}
\date{\today}

\begin{abstract}
One important way to characterize the states having a gauge symmetry
 spontaneously broken over multibands should be to look at their 
collective excitation modes.  
We find that a three-band system has {\it multiple Leggett modes} 
with significantly different masses, which can be classified 
into different dynamical classes according to whether 
multiple inter-band Josephson currents add or cancel.  
This provides a way to dynamically characterize multiband 
superconductivity while the pairing symmetry is 
a static property.
\end{abstract}

\maketitle
{\it Introduction ---} 
Superconductivity and superfluidity, with their spontaneous broken 
gauge symmetry, harbor unexpected fascinations.  
Specifically, the seminal discovery of the iron-based
superconductor\,\cite{Kamihara;Hosono:2008} has kicked off renewed
interests in multiband superconductivity.   
There, the gauge-symmetry breaking involves multiple bands, 
so the ground state is expected to possess the 
features that can occur only  
in multiband systems, such as the $s_{\pm}$ wave pairing 
proposed for the iron-based superconductor\,\cite{Kuroki;Aoki:2009}.   
Even more interesting are excited states, 
especially the collective excitations of condensate phases 
associated with the broken gauge
symmetry\,\cite{Umezawa:1993}.  
For two-band superconductors, there is a classical 1966 work by 
Leggett, who has shown that a two-band superconductor 
accommodates a special collective excitation, now called Leggett's mode, 
emerging as an out-of-phase mode between different
superfluids\,\cite{Leggett:1966}. 
The Leggett's mode was experimentally detected in a two-band
superconductor, $\mbox{MgB}_{2}$ (Ref.\,\cite{Xi:2008}) via the
Raman scattering\,\cite{Blumberg;Karpinski:2007}.  

However, little is known as to what happens to the collective modes 
when there are three or more bands.  
In fact, despite intensive studies on the gap function symmetries 
in the ground states for various classes of superconductors, 
dynamics of collective excitations 
in multiband systems has yet to be systematically investigated.  
The question is becoming increasingly realistic, since 
the iron-based superconductors, for instance, 
have been revealed to have a five-band structure, where 
three bands contribute to both the Fermi surface and 
the gap function\,\cite{Kuroki;Aoki:2009}, 
as subsequently experimentally examined\,\cite{Ishida;Hosono:2009}.  
To study the dynamics of collective 
modes in multiband systems is thus 
expected to be a new avenue together with unconventional pairing
symmetry.  

The purpose of the present paper is to examine, in general, 
whether 
a novel class of Leggett's modes indeed exists specifically 
in superconductors and superfluids that have three or more bands, 
which should shed light on dynamics of multiband condensates.  
Naively, one might expect that the two- and 
three-band cases may be similar.  
Here, however, we shall show, on the basis of an effective action for the
phase fluctuations, that multiband superfluids having three or more
bands are in fact unique in that there exist multiple 
Leggett's modes classified by a dynamical 
class introduced here. 
The distinction between the dynamical classes come from the presence of
multiple inter-band Josephson couplings, whose additive and subtractive
combination to the effective action are, respectively, classified as
class ``even'' and ``odd.''   
The two Leggett's modes  in the class odd are predicted to have 
significantly different masses, which is testable and 
serves to characterize multiband superconductors.  

{\it Formulation ---} 
Let us start with a model for multiband
superfluidity/superconductivity. 
We take the simplest possible Bardeen-Cooper-Schrieffer type
Hamiltonian density for an $N$-band superfluid,  
assumed to be neutral here, is given as     
\(
 \mathcal{\Hop}(\vecr) 
=
\mathcal{\Hop}_{\rm kinetic}(\vecr)
+ \mathcal{\Hop}_{{\rm pairing}}(\vecr)
\) 
with 
\(\mathcal{\Hop}_{\rm kinetic}
=
\sum_{i=1}^{N}\sum_{\sigma=\uparrow,\downarrow}
\psiop^{\dagger}_{i\sigma} 
\varepsilon_{i}(-i\nabla) \psiop_{i\sigma}
\) 
where the dispersion $\varepsilon_{i}(\veck)$ is assumed to be 
parabolic with $\hbar=1$, while 
\(
\mathcal{\Hop}_{\rm pairing}
=
-\sum_{i,j=1}^{N}
g_{ij}
\psiop^{\dagger}_{i\uparrow} \psiop^{\dagger}_{i\downarrow}
\psiop_{j\downarrow} \psiop_{j\uparrow}
\).  
Here $\psiop_{i\sigma}$  ($i=1,\,2,\ldots,N$) 
is the field operator for the $i$th band 
in a superconductor or $i$th atomic species in an atomic gas, 
and $\GVec{\mathcal{G}}\equiv(g_{ij})$ the pairing matrix, taken to be 
positive-definite, and we drop its $\veck$-dependence.   
While $\GVec{\mathcal{G}}$ may come from (the short-range part of) the 
electron correlation, we ignore the Coulombic part 
of the interaction, 
since the mass of Leggett's modes is not affected by the 
Anderson-Higgs mechanism, although 
the group velocity of the mode reflects the Coulomb 
interaction in superconductors as shown in
Ref.\,\cite{Sharapov;Beck:2002} for $N=2$.
While $\Hop_{\rm pairing}$ with $\veck$-dependence 
dropped, as is done in Leggett's original treatment, is too simple to
examine complicated band structures as in the iron-based
superconductors, we should initially know 
the basic physics for collective excitations in multiband
superfluidity/superconductivity in the simplest case. 

The grand partition function 
is given, with the imaginary-time functional integral
method,  in terms of a set of auxiliary fields $\Delta^{(i)}$ as 
\(
 Z 
=
Z_{0} \int 
\prod_{i=1}^{N}\piD \Delta^{(i)} \piD \Delta^{(i)\ast} 
\,e^{-S_{\rm eff}}.
\) 
Here the effective action is 
\(
S_{\rm eff}
=
\int^{\beta}_{0} d\tau 
\int d^{3}\vecr
\sum_{i,j=1}^{N}\Delta^{(i)\ast}
(\GVec{\mathcal{G}}^{-1})_{ij}\Delta^{(j)}
- \sum_{i=1}^{N}\Tr \ln \Gop_{0,i} 
- \sum_{i=1}^{N}\Tr \ln \Gop_{i}^{-1}
\), where we assume the inverse $\GVec{\mathcal{G}}^{-1}$ 
is well-defined (i.e., $\det \GVec{\mathcal{G}}>0$), 
$Z_{0}$ the partition function for the non-interacting system, and 
$\beta$ the inverse temperature. 
$\Gop_{i}$ is 
Green's function in  the Nambu representation  for the 
$i$th-band, which satisfies 
\(
 \Gop^{-1}_{i}
=\Gop^{-1}_{0,i}(\Iop - \Gop_{0,i}\Kop_{i})
\), where $\Gop_{0,i}$ is the free-fermion Green's function, 
$\Iop$ the unit matrix, and 
\(
\Kop_{i} 
= \sigma_{1}{\rm Re}\Delta^{(i)} - i\sigma_{2}{\rm Im}\Delta^{(i)}
\), with $\sigma_{\alpha}$ being the Pauli matrices. 
The gaps are coupled 
through the inverse of the pairing matrix ($\GVec{\mathcal{G}}^{-1}$), and 
we shall see that the non-zero off-diagonal elements of $\GVec{\mathcal{G}}^{-1}$ 
determine the Leggett's modes. 

\begin{figure}[tbp]
\centering
\scalebox{0.24}[0.24]{\includegraphics{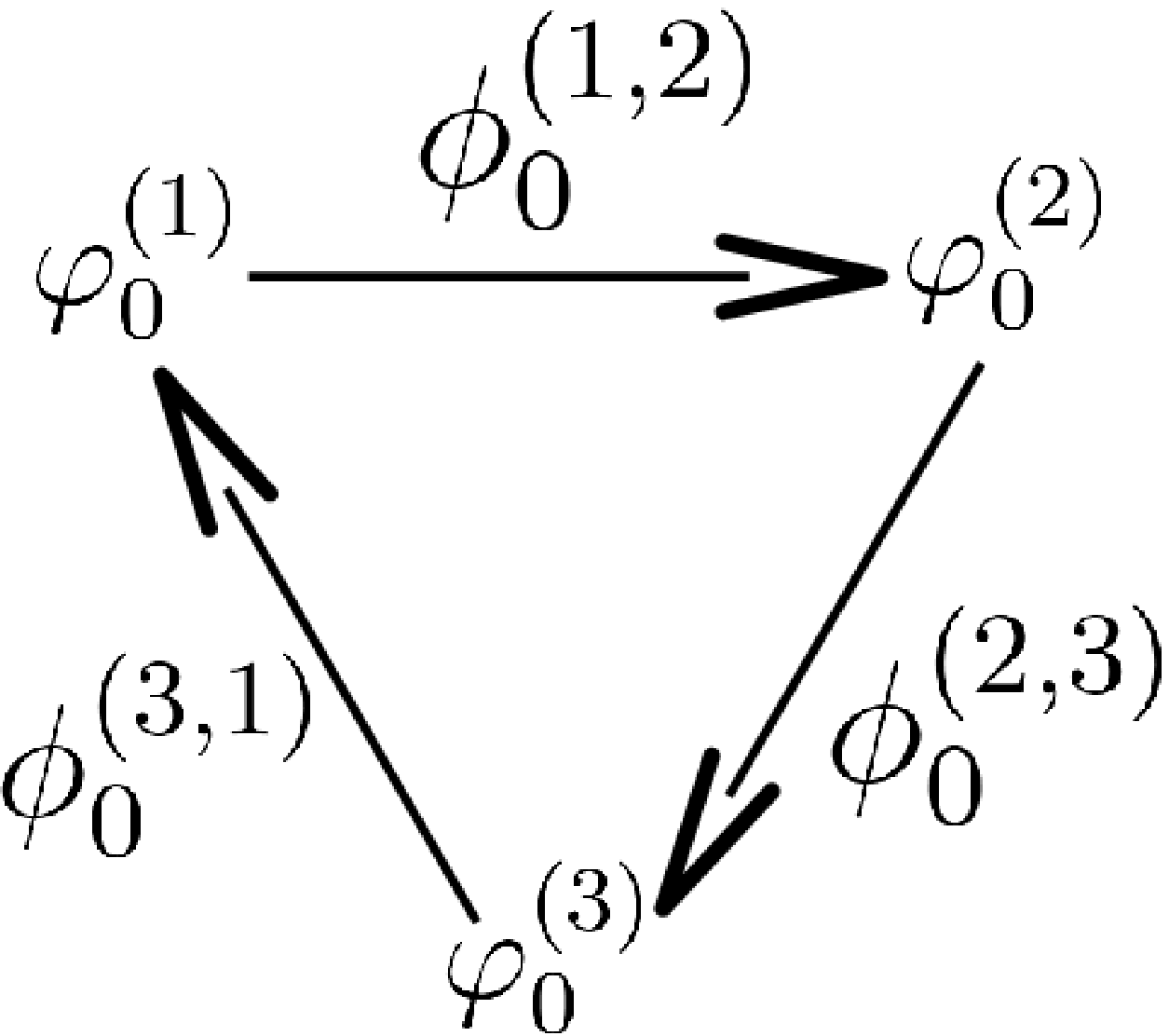}}\,\,
\scalebox{0.24}[0.24]{\includegraphics{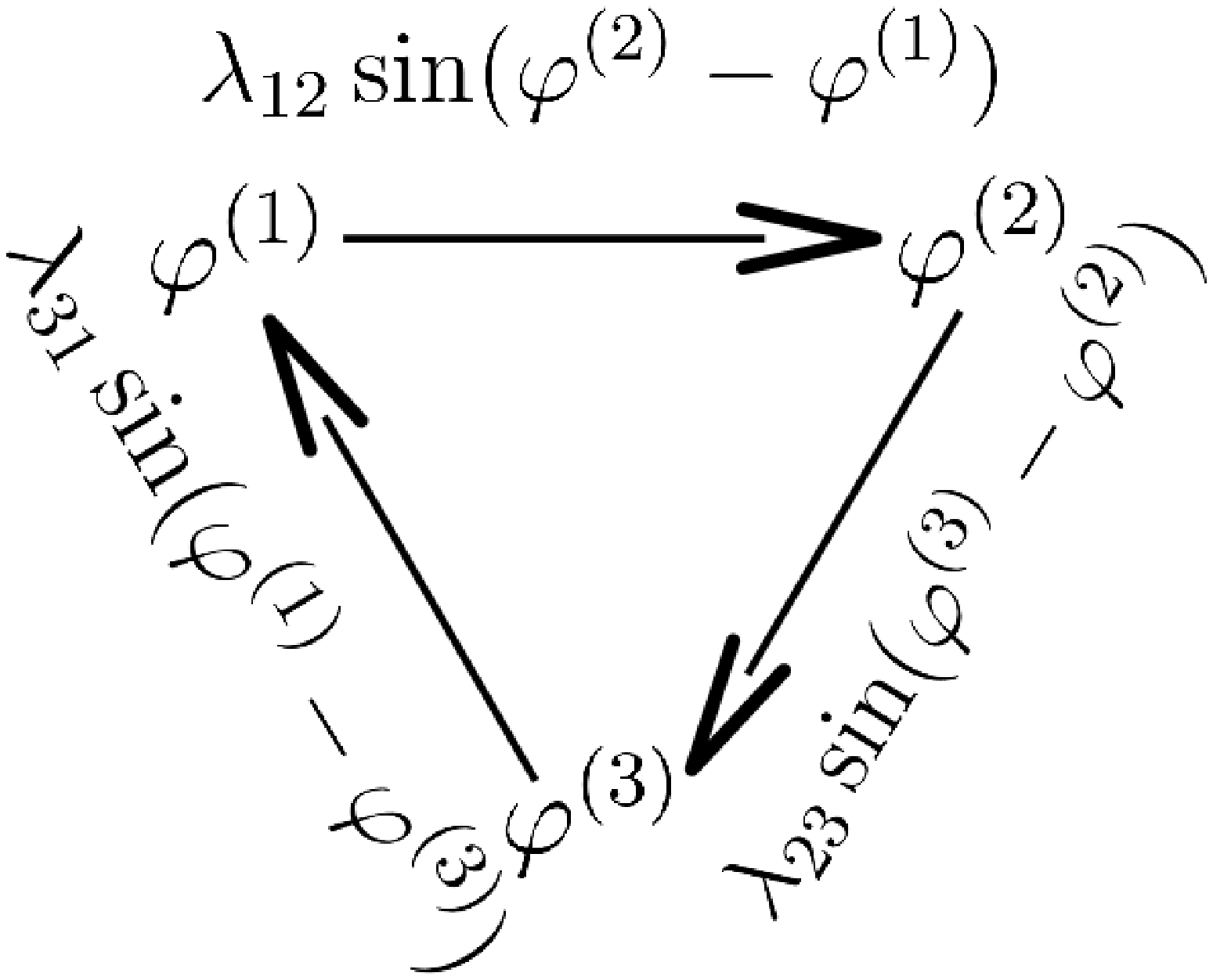}}
\vspace{-2mm}
\begin{flushleft}
(a)\hspace{37mm}(b)
\end{flushleft} 
\caption{
Schematic diagrams for $N=3$ 
for (a) the differences in the superfluid phases 
in equilibrium $\varphi^{(i)}_{0}$ 
and (b) the inter-band Josephson coupling for the phase fluctuation 
$\varphi^{(i)}$ around the equilibrium (where arrows represent the phase 
difference associated with the inter-band Josephson current).}
\label{fig:interJ} 
\end{figure}

With this effective action the static gap
equation reads\,
\begin{equation}
 \Delta_{i}
= 
 \sum_{j=1}^{N} g_{ij}
\Delta_{j}
N_{j}\int^{\omega_{\rm c}}_{0}d\xi 
\frac{\tanh(\beta E_{j}/2)}{E_{j}},
\label{eq:gap_eq}
\end{equation}
where 
\(
E_{i} \equiv \sqrt{\xi^{2} +|\Delta_{i}|^{2}}
\), 
$\omega_{\rm c}$ a cut-off frequency, and $N_{j}$ the
density of states (DOS) of the $j$th fermion on the Fermi surface. 
Here, we assume that each of the $\Delta^{(i)}$'s is constant 
(an $s$-wave).  
We can then look at the phase $\varphi^{(i)}_{0}$ in 
\(
\Delta_{i} = |\Delta_{i}|e^{i\varphi^{(i)}_{0}}
\).  The gap functions on different bands can 
take either the same or opposite signs, 
i.e., the phase difference, 
\(
 \phi^{(i,j)}_{0} \equiv \varphi^{(j)}_{0} 
- \varphi^{(i)}_{0}
\), 
takes either $0$ or $\pm\pi$ 
and should obviously satisfy 
\(
 \sum_{i=1}^{N}\phi_{0}^{(i,i+1)} \equiv 0 
\, \mbox{mod}\,2\pi
\) 
with $\varphi_{0}^{(N+1)}\equiv\varphi_{0}^{(1)}$, as 
depicted in Fig.\ref{fig:interJ}(a).   

Let us derive an effective action for the fluctuations in the 
superfluid phase to single out the collective 
dynamics at zero temperature.   
We first decompose the phase 
\( 
\varphi^{(i)}_{0} + \varphi^{(i)}
\), into the equilibrium $\varphi^{(i)}_{0}$ (as obtained in the mean-field 
gap equation) and the phase fluctuation 
$\varphi^{(i)}$.  
Around the solution of Eq.\,(\ref{eq:gap_eq}), we obtain
an action, 
\(
S_{\rm eff}
=\int^{\beta}_{0}d\tau\int d\vecr V_{\rm interband}
+ \sum_{i=1}^{N}\sum_{m=1}^{\infty}
\Tr(\Gop_{0,i}^{\prime}\Kop_{i}^{\prime})^{m}/m
\), 
where 
\begin{eqnarray}
 V_{\rm interband} 
= \frac{1}{2}\sum_{i<j}
\eta_{ij}\lambda_{ij}
[1-\cos(\varphi^{(j)}-\varphi^{(i)})],
\label{eq:inter}\\
 \eta_{ij}
= \cos(\phi^{(i,j)}_{0}+\kappa_{ij}), 
\quad
\lambda_{ij} = 4|(\GVec{\mathcal{G}}^{-1})_{ij}||\Delta_{i}||\Delta_{j}|,
\label{eq:def_sign}
\end{eqnarray}
comprises a sum of Josephson-couplings 
between the phases of different superfluid gaps  
that represents the 
inter-band Josephson currents caused by the relative phase
fluctuations. 
We mention that the inter-band Josephson couplings are analogous
quantiles with the standard Josephson couplings but derived in
non-perturbative manner. 
In the above expression,  the constraint $\phi_{0}^{(i,j)} =0$ or $\pm\pi$ is 
used, and 
$\kappa_{ij} (=0$ or $\pi$) the sign of
$(\GVec{\mathcal{G}}^{-1})_{ij}$, i.e., 
\(
e^{i\kappa_{ij}}
\equiv -{\rm sgn}(\GVec{\mathcal{G}}^{-1})_{ij}
\).  
The primed quantities are gauge-transformed, i.e., 
\(
\Gop_{0,i}^{\prime}(x;x^{\prime}) 
\equiv 
\Uop_{i}(x)
\Gop_{0,i}(x;x^{\prime})
\Uop_{i}^{\dagger}(x^{\prime}), 
\) 
where 
\(
\Uop_{i} = e^{-i\sigma_{3}(\varphi_{0}^{(i)}+\varphi^{(i)})/2}
\) 
and 
\(
x\equiv (\tau,\vecr)
\), 
with the Dyson equation transformed into
\(
\Gop^{\prime\,-1}_{i}
=\Gop^{\prime\,-1}_{0,i}(\Iop - \Gop^{\prime}_{0,i}\Kop^{\prime}_{i})
\).  

{\it Parity in the inter-band Josephson couplings ---} 
Let us begin with a trivial two-band ($N=2$) case, 
which is relevant to the superconductivity in $\mbox{MgB}_{2}$\,\cite{Xi:2008}.  
Equation (\ref{eq:inter}) then reduces to 
\(
V_{\rm interband}
= (\eta_{12}\lambda_{12}/2)[1-\cos(\varphi^{(2)}-\varphi^{(1)})]
\) with a single inter-band Josephson coupling.  
The sign of $-(\GVec{\mathcal{G}}^{-1})_{12}(=g_{12}/\det \GVec{\mathcal{G}})$ is
equal to that of $g_{12}$. 
Hence $\kappa_{12}=0$ when $g_{12}>0$, 
for which $\phi^{(1,2)}_{0}=0$ 
for $V_{\rm interband}$ to give a
stable gap solution according to Eq.\,(\ref{eq:def_sign}). 
Similarly, $\kappa_{12}=\phi^{(1,2)}_{0}=\pi$
when $g_{12}<0$. 
Thus we always have $\eta_{12}=1$, 
which implies that the sign of $g_{12}$ is totally irrelevant 
to the spectrum of collective modes
in the two-band case, as noted in
Refs.\,\cite{Leggett:1966,Sharapov;Beck:2002,Iskin;SadeMelo:2006}. 

We now turn to the three-band case with $N=3$.  
We have three kinds of the inter-band Josephson currents 
as schematically depicted in Fig.\,\ref{fig:interJ}(b).  
In this case the set of the signs of the couplings 
($\eta_{12},\eta_{23},\eta_{31}$) can be classified into: 
(i) all the signs positive, (ii) two positive, one negative, 
(iii) one positive, two negative, and (iv) all negative. 
We note that, when two or three $\eta_{ij}$'s 
are negative, the Hessian matrices
associated with $V_{\rm interband}$ always have negative eigenvalues 
at $(\varphi^{(1)},\varphi^{(2)},\varphi^{(3)})=(0,0,0)$\,\cite{math_memo}, 
which implies that the solution of Eq.\,(\ref{eq:gap_eq}) is not a stable
minimum of $S_{\rm eff}$. 
We can thus exclude cases (iii) and (iv) [Table
\ref{tab:classification}]. 

\begin{table}[tbp]
\begin{tabular}{clccc}
\hline\hline
class
& $(\eta_{12},\eta_{23},\eta_{31})$ 
& parity of $p_{\vec{\kappa}}$
& parity of $p_{\vec{\phi}_{0}}$ \\\hline
even
& $(1,1,1)$ 
& even 
& even \\
odd
& $(1,1,-1)$
& odd 
& even \\
odd
& $(1,-1,1)$
& odd
& even \\
odd
& $(-1,1,1)$
& odd
& even \\\hline\hline
\end{tabular} 
\caption{Classification of the signs of the coupling for $N=3$.}
\label{tab:classification}
\end{table}

Let us have a closer look at $\{\phi_{0}^{(i,j)}\}$ and
$\{\kappa_{ij}\}$, from which 
$\{\eta_{ij}\}$  in Eq.\,(\ref{eq:def_sign}) 
is determined.  
Note first that $\{\kappa_{ij}\}$ is governed by $\GVec{\mathcal{G}}^{-1}$
rather than $\GVec{\mathcal{G}}$, where, e.g., 
\(
-(\GVec{\mathcal{G}}^{-1})_{12}
=(g_{12}g_{33}-g_{13}g_{23})/\det \GVec{\mathcal{G}}
\) for $N=3$. 
If we introduce a shorthand, 
\(
\vec{\kappa}=(\kappa_{12},\kappa_{23},\kappa_{31}) 
\), we can define an integer, 
\begin{equation}
p_{\vec{\kappa}} 
\equiv \kappa_{12}+\kappa_{23} +\kappa_{31} \quad (\mbox{mod}\,\, 2\pi). 
\end{equation}
Since $\kappa_{ij} =0$ or $\pi$, we have either 
$p_{\vec{\kappa}}/\pi=0$ or $1$, which defines the parity 
``even'' and ``odd'' classes, respectively.  
For the phases 
\(
\vec{\phi}_{0}=(\phi^{(1,2)}_{0}, \phi^{(2,3)}_{0}, \phi^{(3,1)}_{0})
\), 
we have four possible cases, $\vec{\phi}_{0}=(0,0,0)$,
$(0,\pi,\pi)$, $(0,\pi, -\pi)$, or $(0,-\pi,\pi)$ 
as seen in Fig.\,\ref{fig:interJ}(a), 
where we assume $\phi^{(1,2)}_{0} = 0$ without a loss of generality.  
The first case corresponds to an $s$-wave with no sign changes, while
the others sign-reversing $s$-waves.  
If we define 
\(
p_{\vec{\phi}_{0}} \equiv 
\phi^{(1,2)}_{0}+\phi^{(2,3)}_{0}+\phi^{(3,1)}_{0}
\) (mod $2\pi$), we have always 
$p_{\vec{\phi}_{0}}\equiv 0$.  

In the class even, in which all $\eta_{ij}$'s are $1$, 
we find, for all the above possibilities for 
$\vec{\kappa}$ and $\vec{\phi}_{0}$, 
that the class even occurs only when $p_{\vec{\kappa}}\equiv 0$. 
On the other hand, the class odd, where one of
$\eta_{ij}$'s is $-1$, should satisfy 
$p_{\vec{\kappa}}\equiv 1$. 
Thus the class for $\{\eta_{ij}\}$ are 
completely characterized by the parity of $p_{\vec{\kappa}}$ alone, as 
summarized in Table \ref{tab:classification}.  
We remark that $\Tr(\Gop^{\prime}_{i,0}\Kop^{\prime}_{i})^{m}$ 
appearing in $S_{\rm eff}$ does {\it not}
depend on $\{\eta_{ij}\}$, which implies that 
the effective action is completely 
distinguished by the dynamical class characterized by the parity
of $p_{\vec{\kappa}}$.  
Such dynamical classification should always 
be applicable in the pairing-interaction parameter space 
when $N\ge 3$.  
In other words, 
$\{\kappa_{ij}\}$, which does not 
have to coincide with $\{\phi^{(i,j)}_{0}\}$ modulo $2\pi$, 
is {\it no longer} a matter of convention 
unlike the case of $N=2$.  

\begin{figure}[tbp]
\centering
\scalebox{0.68}[0.68]{\includegraphics{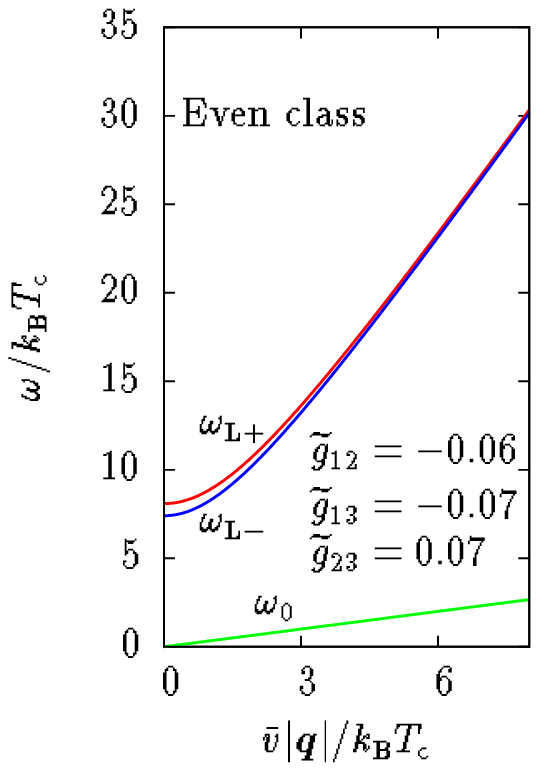}}\hspace{4mm}
\scalebox{0.68}[0.68]{\includegraphics{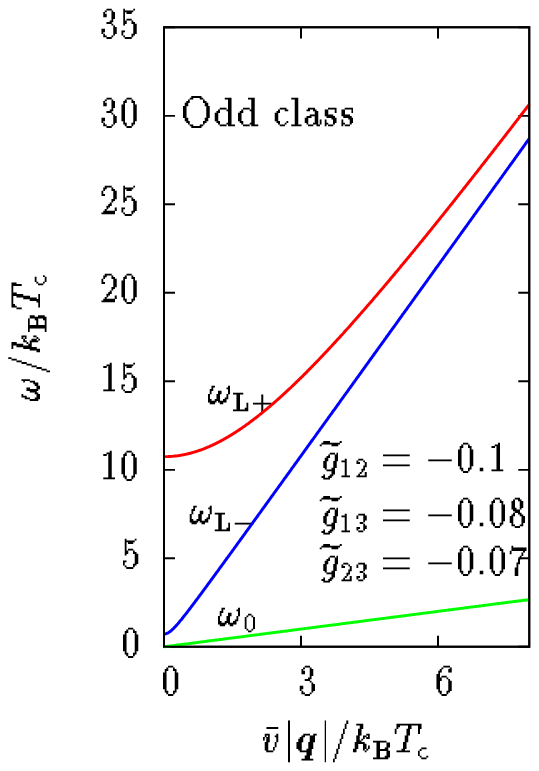}} 
\vspace{-9mm}
\begin{flushleft}
(a)\hspace{40mm}(b) 
\end{flushleft}
\caption{(Color online) Dispersion relations for $\omega_{{\rm L}+}$
 (red line), $\omega_{{\rm  L}-}$ (blue) and $\omega_{0}$ 
(green) for (a) the class even 
with the values of $\tilde{g}_{ij}$ displayed for which 
$\Delta_2/\Delta_1=-1.00$ and $\Delta_3/\Delta_1=-1.01$, and
 (b) the class odd  with 
$\Delta_2/\Delta_1=-0.73$, $\Delta_3/\Delta_1=-0.38$.  
We set $\omega_{\rm c}/k_{\rm B}T_{\rm c}=5000$ and 
$\widetilde{g}_{ii}=\frac{1}{3}$.  
The DOS ($N_i$) and the Fermi velocity are assumed
 to be identical between the bands. }
\label{fig:dispersion}
\end{figure}

{\it Collective modes ---} 
We are now in position to calculate the collective modes.  
If we expand $S_{\rm eff}$ around the stable solution of
Eq.\,(\ref{eq:gap_eq}) 
for $\varphi^{(i)}(q)$ with 
\(
q\equiv (i\nu_{\ell},\vecq)
\) and  
\(
\nu_{\ell}=2\ell\pi/\beta
\) ($\ell\in\mathbb{Z}$) the Matsubara frequency, 
we have, to the leading order, 
\(
S_{\rm eff} 
\approx 
\sum_{q} \!^{t}\!\vec{\varphi}_{q} (\mathcal{M}_{q}/4)\vec{\varphi}_{q}
\) 
in the long-wavelength limit 
($|\vecq|\to 0$)\,\cite{Sharapov;Beck:2002}. 
Here we have introduced 
\(
\vec{\varphi}_{q} 
= \!^{t}(\varphi^{(1)}(q),\varphi^{(2)}(q),\varphi^{(3)}(q))
\), and a $3\times 3$ real symmetric matrix, 
\begin{equation*}
\mathcal{M}_{q}=
\left(
\begin{array}{ccc}
M_{11}(q)+\mu_{11} 
& -\eta_{12}\lambda_{12}
& -\eta_{31}\lambda_{31}\\
-\eta_{12}\lambda_{12}
& M_{22}(q)+\mu_{22}
& -\eta_{23}\lambda_{23}\\
-\eta_{31}\lambda_{31}
& -\eta_{23}\lambda_{23}
& M_{33}(q)+\mu_{33} \\
\end{array}
\right),
\end{equation*}
where  
\(
M_{ii}(q)
=N_{i}\nu_{\ell}^{2} + \sum_{\xi}(N_{i}v_{{\rm F}\xi,i}^{2}/3)q_{\xi}^{2}
\) 
with 
\(
v_{{\rm F}\xi,i} 
\) the $\xi$th component ($\xi=x,y,z$) of the $i$th fermion's Fermi 
velocity. 
The diagonal elements involve the contributions from
the inter-band Josephson currents, 
\(
\mu_{11}=\eta_{12}\lambda_{12}+\eta_{31}\lambda_{31}
\), etc.
The dispersion relations for the collective excitation
modes are given by the roots of $\det\mathcal{M}_{q}=0$. 
After an analytic continuation $i\nu_{\ell}\to \omega + i0$,
we obtain the three roots, $\omega_{0}$, $\omega_{{\rm L}+}$, and
$\omega_{{\rm L}-}$. 
Figure \ref{fig:dispersion} displays the full dispersion.  
We can give explicit formulae for the
three collective modes to the leading order in $q$ for isotropic
systems. 
The first root $\omega_{0}$ corresponds to the Nambu-Goldstone 
mode with 
\( 
\omega_{0} = V |\vecq| + O(|\vecq|^{2}) 
\), where 
\(
V^{2} = 
\sum_{i=1}^{3}(N_{i}/N_{\rm tot})(v_{{\rm F}i}^{2}/3)
\) 
and 
\(N_{\rm tot}=\sum_{i}N_{i}\). 
The other two roots, $\omega_{{\rm L}+}$ and $\omega_{{\rm L}-}$, 
are the Leggett's modes
\(
\omega_{{\rm L}\pm}^{2} 
= \nu_{\pm}^{2} + c_{\pm}^{2}|\vecq|^{2} + O(|\vecq|^{4})
\), where 
\(
c_{\pm}^{2} 
= 
\pm U^{2}
\mp \nu_{\mp}^{2}V^{\prime\,2}/(\nu_{+}^{2}-\nu_{-}^{2})
\). 
Here the ``mass gap'' (the Leggett's mode frequency at $|\vecq|=0$, 
i.e., $\nu_{+}, \nu_{-}$) are characterized by two quantities: 
The frequency scale is given by  
\(
 \bar{\nu}^{2}
=\eta_{12}\nu_{12}^{2}
 + \eta_{23}\nu_{23}^{2}+\eta_{31}\nu_{31}^{2}
\) and 
\(
\nu_{ij}^{2}
= (N_{i}+N_{j})\lambda_{ij}/N_{i}N_{j}
\), 
where $\nu_{ij}$ for $\lambda_{ij}\neq 0$ corresponds to the frequency
of the Leggett's mode in the two-band model. 
The difference $\nu_{+}-\nu_{-} (\propto \sqrt{1-D})$ is characterized by
\begin{equation*}
D
=
\frac{4N_{\rm tot}(
\eta_{12}\eta_{31}\lambda_{12}\lambda_{31}
+
\eta_{12}\eta_{23}\lambda_{12}\lambda_{23}
+
\eta_{23}\eta_{31}\lambda_{23}\lambda_{31}
)}{\bar{\nu}^{4}N_{1}N_{2}N_{3}}.
\end{equation*}
For $D\to 1$ the mass difference $\nu_{+}-\nu_{-}$ vanishes, whereas 
for $D\to 0$ $\omega_{{\rm L}-}$ becomes a gapless mode (i.e., 
\(
\omega_{{\rm L}-}\to c_{-}|\vecq|
\)).  
We find that $\nu_{+}-\nu_{-}$ becomes large due to 
a cancellation among the terms in the numerator of $D$ 
occurs for class odd.  
The term quadratic in $|\vecq|$ for the two Leggett's modes 
are characterized by $c_{\pm}$, which involve two velocities,  
\(
U^{2}
=
(1-D)^{-1/2}\sum_{i=1}^{3}
(\mu_{ii}/N_{i}\bar{\nu}^{2}) 
(v^{2}_{{\rm F}i}/3)
\)
and 
\(
V^{\prime 2}
=
\bar{v}^{2}/3- V^{2}
\),
with 
\(
\bar{v}^{2} = \sum_{i=1}^{3}v_{{\rm F}i}^{2}
\). 
In the limit where two bands are decoupled 
($\lambda_{23}=\lambda_{31}=0$), we 
recover the two-band result\,\cite{Sharapov;Beck:2002} 
with two gapless modes ($\omega_{0}, \omega_{{\rm L}-}$), and one
gapped mode ($\omega_{{\rm L}+}$), along with 
$\nu_{+}$ and $\nu_{-}$ that coincide with those in
Ref.\,\cite{Kochorbe;Palistrant:1998}. 

\begin{figure}[tbp]
\scalebox{0.72}[0.72]{\includegraphics{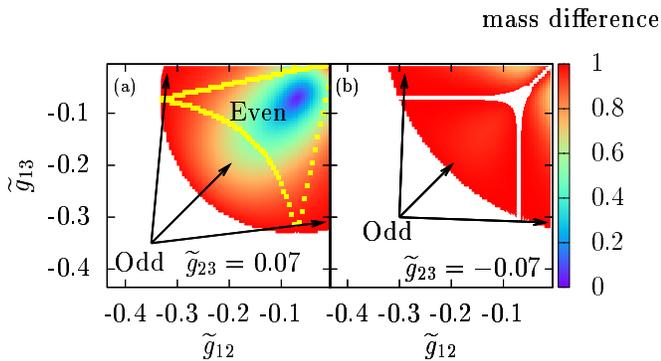}}
\caption{(Color online) Color-coded mass difference 
 ($\propto \sqrt{1-D}$) in the multiple Leggett's
 modes plotted on $\widetilde{g}_{12}$-$\widetilde{g}_{31}$ plane for
 $\widetilde{g}_{ii}=\frac{1}{3}$.  
We assume identical normal state properties between the three bands.  
(a) Case of two negative (repulsive) and one positive (attractive)
 inter-band couplings. Yellow lines represent the boundaries between the
 even and odd classes. (b) Case of all the three  inter-band couplings
 negative (repulsive), where only the class odd is allowed. In the blank
 area, the Leggett's modes become unstable.}
\label{fig:profile}
\end{figure}

Having formulated the Leggett's modes, let us see how they reflect the
difference in the dynamical classes even and odd. 
If we look at the simplest case of 
\(
\widetilde{N}_{i}\equiv N_{i}/N_{\rm tot}=\frac{1}{3},
\) 
and 
\(
\widetilde{v}_{{\rm F}i}^{2}\equiv 
v_{{\rm F}i}^{2}/\bar{v}^{2}=\frac{1}{3}
\) 
(for $i=1,2,3$), 
for  
\(
\widetilde{g}_{11}=\widetilde{g}_{22}=\widetilde{g}_{33}
=\frac{1}{3}
\)
with 
\(
\widetilde{g}_{ij}\equiv g_{ij}/\Tr \GVec{\mathcal{G}}
\), 
Fig.\ref{fig:dispersion} displays typical dispersion relations for
$\omega_{0}$ and $\omega_{{\rm L}\pm}$ for the two dynamical classes. 
We immediately notice that the mass difference 
$\nu_{+}-\nu_{-}(=\omega_{{\rm L}+}-\omega_{{\rm -}}$ at $|\vecq|=0$)
is much greater in the class odd than in the class even.  
Systematic variation of  $\nu_{+}-\nu_{-}$ on an interaction parameter
space ($\widetilde{g}_{12}$,$\widetilde{g}_{13}$) is shown, 
first  in Fig.\,\ref{fig:profile}(a) 
for the case of two negative (repulsive) and one positive (attractive) 
inter-band couplings. 
The allowed parameter region is restricted by the positive-definiteness
of $\GVec{\mathcal{G}}$\,\cite{Kimura:2003}. 
The mass difference is close to zero in the central area of the class
even region, while it is relatively large in the class odd region. 
Thus, we recognize that the mass difference of the Leggett's modes is an 
indicator of the difference in the dynamical classes.  
Note, however, that the distinction becomes blurred 
when the three-band system approaches a two-band behavior, 
which occurs when one of the
off-diagonal elements of $\GVec{\mathcal{G}}^{-1}$ is much greater than the others
(e.g., around the corners of the class odd region with a
triangular shape), so that there is no jumps at the boundary. 

Next, Fig.\,\ref{fig:profile}(b) displays the result for the case of
all the three  inter-band couplings negative (repulsive)
with $g_{12},g_{13},g_{23}<0$.  Notably, 
this case always has the class odd throughout. 
This implies that the mass difference
of the Leggett's modes is  {\it lower bounded}. 
The result shows that the bound is $\simeq 0.7$ for
$\widetilde{g}_{23}=-0.07$ and $g_{ii}=\frac{1}{3}$.  
Even more interesting, we find 
that the Leggett's modes become unstable or ill-defined 
(within the present treatment which assumes real $\Delta_i$'s) 
in a narrow but finite region (blank area in 
Fig.\,\ref{fig:profile}(b)) where all three (or two) of $g_{12},g_{13},g_{23}$ 
take similar values.  
In this case, every two among 
the three gap functions $\Delta_1,\Delta_2,\Delta_3$  in 
Eq.\,(\ref{eq:gap_eq})
want to have opposite signs, but end up with 
complex values (i.e., the relative phase differences
between the gaps deviate from $0$ or $\pi$) 
due to the ``frustration'', as pointed
out in Ref.\,\cite{Stanev;Tesanovic:2010}. 

{\it Summary ---}
We have shown the presence of multiple dynamical classes in the $N(\ge
3)$-band superfluidity, which are characterized in terms of the parity
of the multiple inter-band Josephson couplings. 
We have revealed that the mass difference of the Leggett's modes
is greater in the class odd, in
contrast to the two-band case where the classification does not exist. 
So, the behavior of the multiple Leggett's modes is expected to
characterize the dynamics of excitations in multiband superconductors
and superfluids. 
The expressions for the parities for $N=3$ given here can be extended to
$N\ge 4$, with their classification being an interesting future problem.  
Other future works include an extension of the present discussion to
more general, $\veck$-dependent case for realistic description of
multiband superconductors, such as the iron-based superconductors with
their material dependence\cite{Kuroki;Aoki:2009}, and an extension to
anisotropic pairings. 
Another intriguing problem is the Leggett's modes in the case 
where the gap functions break the time-reversal symmetry 
in ``frustrated'' three-band
superconductivity\,\cite{Stanev;Tesanovic:2010}. 
Multiband superfluidity in cold fermionic atomic gases with multiple
hyperfine states may also be an interesting playing 
ground where we have a greater tunability 
due to the Feshbach resonance\cite{Iskin;SadeMelo:2006}.   
How to detect Leggett's mode, for which Raman spectra have been
discussed in
Ref.\,\cite{Chubukov;Korshunov:2009,Scalapino;Devereaux:2009,Klein:2010,Burnell;Bernevig:2010},
is also an important problem. 

YO and MM wish to thank illuminating discussions with Y. Yamanaka,
H. Nakamura, M. Okumura, N. Nakai, Y. Nagai, and M. Mine. 
HA acknowledges valuable discussions 
with A. Leggett, K. Kuroki and R. Arita.


\begin{thebibliography}{99}
\bibitem{Kamihara;Hosono:2008}
Y. Kamihara, T. Watanabe, M. Hirano, and H. Hosono, 
J. Am. Chem. Soc. {\bf 130}, 3296 (2008). 
\bibitem{Kuroki;Aoki:2009}
K. Kuroki {\it et al.}, 
Phys. Rev. Lett. {\bf 101}, 087004 (2008); 
{\bf 102}, 109902(E) (2009); 
K. Kuroki {\it et al.}, 
Phys. Rev. B {\bf 79}, 224511 (2009). 
\bibitem{Umezawa:1993}
H. Umezawa, 
{\it Advanced Field Theory: Micro, Macro, and Thermal Physics} 
(AIP press, New York, 1993). 
\bibitem{Leggett:1966}
A. J. Leggett, 
Prog. Theor. Phys. {\bf 36}, 901 (1966).
\bibitem{Xi:2008}
X. X. Xi, 
Rep. Prog. Phys. {\bf 71}, 116501 (2008).
\bibitem{Blumberg;Karpinski:2007}
G. Blumberg {\it et al.}, 
Phys. Rev. Lett. {\bf 99}, 227002 (2007).
\bibitem{Ishida;Hosono:2009}
K. Ishida, Y. Nakai, and H. Hosono, 
J. Phys. Soc. Jpn. {\bf 78}, 062001 (2009). 
\bibitem{Sharapov;Beck:2002}
S. G. Sharapov, V. P. Gusynin, and H. Beck, 
Eur. Phys. J. B {\bf 30}, 45 (2002). 
\bibitem{Iskin;SadeMelo:2006}
M. Iskin and C. A. R. S$\acute{\rm a}$ de Melo, 
Phys. Rev. B {\bf 74}, 144517 (2006).  
\bibitem{math_memo}
If a hermitian matrix has a negative and non-zero
diagonal element, it has a
negative and non-zero eigenvalue.  
\bibitem{Kochorbe;Palistrant:1998}
F. G. Kochorbe and M. E. Palistrant, 
Physica C {\bf 298}, 217 (1998). 
\bibitem{Kimura:2003}
G. Kimura, 
Phys. Lett. A {\bf 314}, 339 (2003). 
\bibitem{Stanev;Tesanovic:2010}
V. Stanev and Z. Te$\check{\rm s}$anovi$\acute{\rm c}$, 
Phys. Rev. B {\bf 81}, 134522 (2010).
\bibitem{Chubukov;Korshunov:2009}
A. V. Chubukov, I. Eremin, and M. M. Korshunov, 
Phys. Rev. B {\bf 79}, 220501(R) (2009).
\bibitem{Scalapino;Devereaux:2009}
D. J. Scalapino and T. P. Devereaux, 
Phys. Rev. B {\bf 80}, 140512(R) (2009).
\bibitem{Klein:2010}
M. V. Klein, 
Phys. Rev. B {\bf 82}, 014507 (2010).
\bibitem{Burnell;Bernevig:2010}
F. J. Burnell {\it et al.}, 
Phys. Rev. B {\bf 82}, 144506 (2010).
\end{thebibliography}
\end{document}